\documentclass{ws-ijmpcs}

\begin{document}
\markboth{A. Sulaiman and L.T Handoko}
{Modeling the interaction of biomatter and biofluid}

%
\catchline{}{}{}{}{}
%
\def\grad{\mathbf{\nabla}}
\def\om{\mathbf{\omega}}
\def\v{\mathbf{v}}
\def\x{\mathbf{x}}
\def\u{\mathbf{U}}
\def\f{\mathbf{F}}
\def\s{\mathbf{S}}
\def\en{\mathbf{E}}
\def\r{\mathbf{r}}
\def\mb{\bar{m}}
\def\mt{\tilde{m}}
\def\lt{\tilde{\lambda}}
\def\xp{{x^\prime}}
\def\l{{\cal L}}
\def\cd{{\cal D}}
\def\h{{\cal H}}
\def\f{{\cal F}}
\def\z{{\cal Z}}
\def\s{{\cal S}}
\def\e{{\cal E}}
\def\p{{\cal P}}
\def\cd{{\cal D}}
\def\cj{{\cal J}}
\def\pd{\partial}
\def\d{\mathrm{d}}
\def\j{\mathbf{J}}
\def\jt{\tilde{J}}
\def\lt{\tilde{\lambda}}
\def\mt{\tilde{m}}
\def\at{\tilde{\alpha}}
\def\jtv{\mathbf{\tilde{J}}}
\def\exp{\mathrm{exp}}
\def\ex{\mathrm{e}}
\def\be{\begin{equation}}
\def\ee{\end{equation}}
\def\bea{\begin{eqnarray}}
\def\eea{\end{eqnarray}}
\def\ie{\textit{i.e.} }

\title{\bf MODELING THE INTERACTIONS OF BIOMATTER AND BIOFLUID }

\author{A. SULAIMAN}
\address{Badan Pengkajian dan Penerapan Teknologi, BPPT Bld. II (19$^{\rm th}$
floor), Jl. M.H. Thamrin 8, Jakarta 10340, Indonesia\\
asulaiman@webmail.bppt.go.id, sulaiman@teori.fisika.lipi.go.id}

\author{L.T. HANDOKO}
\address{Group for Theoretical and Computational Physics,
Research Center for Physics, Indonesian Institute of Sciences,
Kompleks Puspiptek Serpong, Tangerang, Indonesia\\
handoko@teori.fisika.lipi.go.id, handoko@fisika.ui.ac.id,
laksana.tri.handoko@lipi.go.id}
\address{Department of Physics, University of Indonesia,
Kampus UI Depok, Depok 16424, Indonesia}

\maketitle

\begin{history}
\received{Day Month Year}
\revised{Day Month Year}
\end{history}

\begin{abstract}
The internal motions of biomatter immersed in biofluid are
investigated. The interactions between the fragments of biomatter
and its surrounding biofluid are modeled using field theory. In
the model, the biomatter is coupled to the gauge field
representing the biofluid. It is shown that at non-relativistic
limit various equation of motions, from the well-known Sine-Gordon
equation to the simultaneous nonlinear equations, can be
reproduced within a single framework.

\keywords{biomatter; biofluid; soliton.}
\end{abstract}

\ccode{PACS numbers: 11.25.Hf, HAKLUYT}

\section{Introduction}

The deoxyribonucleic acid (DNA) is getting the very important
biomolecules. Especially its helical structures undergoes a very
complex dynamics and plays several important roles in various
biological phenomena such as storage of information, inheritance
(replication, etc) and the usage of genetic information
(transcription, etc). Physically, transcription processes begin with the
open state in the DNA double helix. The phenomena describe  the internal DNA mobility
such as rotational motion of nitrous bases, asymmetry of the helix, strength of
the hydrogen bond and so on \cite{yakushevich}. The rotational motion mode with
such asymmetry has a solitary wave solution describing the open state of double
helix \cite{yakushevich2}.

DNA is not motionless. It is in a constantly wriggling dynamics state in a
medium of bio-organic fluid in the nucleus cell. One can think of the DNA
molecule as behaving somewhat like a microscopic piece of worm-like elastic
suspended in solution. In this approximation, one ignores all details of the
chemical structure. The motion of DNA surrounded by fluid is rarely studied
\cite{yakushevich,voet}. Previous studies are usually done by
solving the fluid equations and its wave equations simultaneously
using appropriate boundary conditions \cite{yakushevich}. For 
example, the thermal denaturation of double strand DNA depends on
the solution surrounding the DNA molecules. It is necessary to take
into consideration that the solving water does act as a viscous
medium and may damp out bubble denaturation. The impact of
viscosity has been investigated by Zdrakovic et al \cite{kovici} which shows 
that the behavior of DNA dynamics in viscous solution is
described by the damped nonlinear Schrodinger (NLS) equation.  The
solution is then obtained by expansion and performing
order-by-order calculation. In these approaches, anyway the
picture of interaction between DNA and its surrounding fluid is
not clear. Also, in most models the over-damped DNA dynamics are
treated by putting some additional terms by hand in the
differential equation to obtain the non-homogeneous ones.

There is a technology for DNA manipulation developed in the last decade using 
gel electrophoresis \cite{smith}. This technique is used to estimate the size
of genetic sequences and understand the dynamics of individual strands of DNA
as they move through a fluid solution. In theoretical polymer dynamics, the
study is mostly focused on describing single-polymer dynamics to improve
models of bulk properties. All of them attempt to understand the behavior
of individual polymers under controlled fluid flows through the Navier-Stokes
equation coupled with elastic strings. Theoretical studies of this matter
is still an open problem, especially related to reological problem.

In this paper, we study the soliton excitation in DNA with
surrounding fluid based on the gauge filed theory formulation
\cite{sulaiman}. The DNA dynamics is modeled as the result of
interactions among matters in a fluid medium using the relativistic and gauge
invariant fluid dynamics lagrangian. The theory is a relativistic one, but we
can take its non-relativistic limit at the final stage to deal with problems in
biomatter as done in some previous works, like in some models using the
ideal gas approximation, Sine-Gordon models etc \cite{yakushevich}. The
lagrangian is originally devoted to model the magnetofluid in plasma as a
relativistic fluid system inspired by the similarity between the dynamical
properties of fluid and electromagnetic field  \cite{handoko}.
Within the model, the interactions of DNA and fluid dynamics are described in a
general way as the results of interactions among the fluid and matter fields.

The paper is organized as follows. In the first chapter the model of
interaction between biomolecule and fluid is given. It is then applied to
describe the dynamics of worm-like DNA molecule with varying velocity. The
paper is finally ended by a summary.

\section{The model}

Some mechanical models of  DNA have been proposed over the years which are 
focused on different biological, physical and chemical processes in which DNA is
involved. One of the important model related to the identification of the
unwinding of double helix in a "bubble" is called the torsional $Y$ model
\cite{yakushevich}. The lagrangian describing the bubble of $Y$ model
is represent by a scalar (boson) field governed by the bosonic lagrangian
\cite{yakushevich,sulaiman},
\be
        \l_\mathrm{matter} = \left( \pd_\mu \Phi \right)^\dagger  \left( \pd^\mu \Phi \right) + V(\Phi) \; ,
        \label{eq:lphi}
\ee where $V(\Phi)$ is the potential. For example in the typical
$\Phi^4-$theory,
\be
    V(\Phi) = \frac{1}{2} m_\Phi^2 \, \Phi^\dagger \Phi - \frac{1}{4!} \lambda \, (\Phi^\dagger \Phi)^2 \; ,
    \label{eq:v}
\ee 
where $m_\Phi$ and $\lambda$ are the mass of matter and the nonlinear constant
\cite{yakushevich}. The hermite conjugate is $\Phi^\dagger \equiv
{(\Phi^\ast)}^T$ for a general complex field $\Phi$ describing a rotational
(torsional motion) of DNA.

The interaction in the $Y$ model with fluid medium is described in the gauge
field theory framework. In this theory, the above bosonic lagrangian is
imposed to be gauge invariant under local (in general non-Abelian) gauge
transformation, $U \equiv \exp[-i T^a \theta^a(x)] \approx 1 - i
T^a \theta^a(x)$ with $\theta^a \ll 1$ \cite{sulaiman}. $T^a$'s are generators
belong to a particular Lie group and satisfy certain commutation
relation $[T^a,T^b] = i f^{abc} T^c$ with $f^{abc}$ is the
anti-symmetric structure constant. The matter field is then
transformed as $\Phi \stackrel{U}{\longrightarrow} \Phi^\prime
\equiv \exp[-i T^a \theta^a(x)] \, \Phi$, with $T^a$ are $n \times
n$ matrices while $\Phi$ is an $n \times 1$ multiplet containing
$n$ elements.  It is well-known that the symmetry in Eq.
(\ref{eq:lphi}) is revealed by introducing gauge fields $A_\mu^a$
which are transformed as $U^a_\mu \stackrel{U}{\longrightarrow}
{U^a_\mu}^\prime \equiv U^a_\mu - \frac{1}{g}  (\pd_\mu \theta^a)
+ f^{abc} \theta^b U^c_\mu$, and replacing the derivative with the
covariant one, $\cd_\mu \equiv \pd_\mu + i g \, T^a U^a_\mu$. The
fluids is represent by the gauge fields that guarantees the
invariant properties of the system. Then the total lagrangian with
some additional terms to keep its gauge invariance is,
\be
    \l = \l_\mathrm{matter} + \l_\mathrm{gauge} + \l_\mathrm{int} \; ,
        \label{eq:l}
\ee where, \bea
        \l_\mathrm{gauge} & = & -\frac{1}{4} S^a_{\mu\nu} {S^a}^{\mu\nu} \; ,
        \label{eq:la} \\
    \l_\mathrm{int} & = & -g J^a_\mu {U^a}^\mu + g^2 \left( \Phi^\dagger T^a T^b \Phi \right) U_\mu^a {U^b}^\mu \; .
        \label{eq:li}
\eea The strength tensor is $S^a_{\mu\nu} \equiv \pd_\mu U^a_\nu -
\pd_\nu U^a_\mu + g f^{abc} U^b_\mu U^c_\nu$, while the 4-vector
current is,
\be
        J^a_\mu = -i \left[ (\pd_\mu \Phi)^\dagger T^a \Phi - \Phi^\dagger T^a (\pd_\mu \Phi) \right] \; .
        \label{eq:j}
\ee The coupling constant $g$ then represents the interaction
strength between gauge field and matter. We should note that,
however the current conservation is realized by the covariant
current $\pd^\mu \cj^a_\mu = 0$ with $\cj^a_\mu \equiv -i \left[
(\pd_\mu \Phi)^\dagger T^a \Phi - \Phi^\dagger T^a (\pd_\mu \Phi)
\right]$.

The gauge boson $U_\mu$ is  interpreted as a ``fluid field'' with
velocity $u_\mu$, and takes the form \cite{sulaiman},
\be
    U^a_\mu = \left( U_0^a, \u^a \right) \equiv u^a_\mu \, \phi \; ,
    \label{eq:a}
\ee with,
\be
        u^a_\mu \equiv \gamma^a (1, -\v^a) \; ,
        \label{eq:ae}
\ee where $\phi$ is an auxiliary boson field, while $\gamma^a
\equiv \left( 1 - |\v^a|^2 \right)^{-1/2}$. Here we adopt the
natural unit, \ie the light speed $c = 1$. We should remark that
the auxiliary field $\phi$ is introduced to keep correct
dimension.

For microfluid one usually works with a single fluid. From the gauge theory
point of view, this is realized by the Abelian gauge lagrangian. The total
lagrangian in this case becomes,
\bea
    \l & = & \left( \pd_\mu \Phi^\ast \right)  \left( \pd^\mu \Phi \right) + \frac{1}{2} m_\Phi^2 \, \Phi^\ast \Phi - \frac{1}{4!}\lambda \,
    \left(\Phi^\ast \Phi \right)^2
    + g^2 \, U_\mu U^\mu \, \Phi^\ast \Phi
    \nonumber \\
    & & - \frac{1}{4} \left( \pd_\mu U_\nu - \pd_\nu U_\mu\right)\left( \pd^\mu {U}^\nu - \pd^\nu {U}^\mu\right)
    + i g \, U^\mu\ \left[ \left( \partial_\mu \Phi^\ast\right) \Phi - \Phi^\ast \left( \partial_\mu \Phi \right)\right] \; ,
    \label{eq:lu1}
\eea 
using Eqs. (\ref{eq:v}), (\ref{eq:l})$\sim$(\ref{eq:li}) and
(\ref{eq:j}). The strength tensor is given by $S^a_{\mu\nu} \equiv \pd_\mu
U^a_\nu - \pd_\nu U^a_\mu$. 

The non-relativistic limit can be obtained by performing a transformation $t
\rightarrow \tau \equiv i t$ and  putting $\gamma \rightarrow 1$ respectively.
Imposing the variational principle of action, one reaches at the Euler-Lagrange
equation \cite{ryder}. Therefore, the equation of motion (EOM) for $\Phi$ reads,
\be
    \left( \partial^2 -  m_\Phi^2 - 2 g^2 \, |\v|^2 \right) \Phi + \frac{1}{3!} \lambda \, \Phi^3 = 0 \; .
    \label{eq:eomu1}
\ee
for a real $\Phi$ field. This result leads to the well-known
nonlinear Klein-Gordon equation. On the other hand, the EOM of $U_\mu$ is
given by,
\be
  \frac{\pd \v}{\pd t} + \left( \v.\nabla \right) \v =-g F\; ,
  \label{eq:fluid}
\ee
where $\v$ is fluid velocity in non-relativistic limit. $F$ is total external
force (see \cite{handoko} for detail derivation). This is the model the
underlying model throughout this paper.

\section{DNA dynamics in the surrounding fluid}

Behavior of the DNA interaction with surrounding fluid depends on the boundary
condition of fluid phenomena. One should deal with the boundary value problem 
for the fluid equation and then solve the DNA dynamics
through Eq.(\ref{eq:eomu1}). For the sake of simplicity, however let us
fix the fluid flow in order to avoid solving the boundary value problem.

\subsection{Constant Velocity}

First of all, consider $\v=v_0$ is  a constant and write
down Eq. (\ref{eq:eomu1}) in the standard form,
\be
  \frac{\pd^2 \Phi}{\pd t^2}-\frac{\pd^2 \Phi}{\pd x^2}
  -\mb_\Phi^2 \Phi + \lambda \Phi^3 =0 \; ,
  \label{eq:nkg1}
\ee
 in term of coordinate $(t,x)$, where $\mb_\Phi^2 \equiv m_\Phi^2 + 2 g^2 \,
v_0^2$. It's well known that the solution of the equation can be represented by
the Jacobi elliptic function \cite{fred}.

For $\lambda > 0$ the solution is,
\be
  \Phi(x,t)=A \mb_\Phi sn\left[ \beta (x+x_0-Ct),m\right] \; ,
  \label{eq:nkg2}
\ee 
where $\beta=\sqrt{\lambda \mb_\Phi^2/((1-C^2)(1+m))}$ ,
$A=\sqrt{2m/(1+m)}$, $C$ is the phase velocity and $m$ is a modulus. In the
limit $m\rightarrow 1$, one obtains the usual kink solitary waves as follow,
\be
  \Phi(x,t)= \mb_\Phi^2 \tanh \left[ \sqrt{\frac{\lambda
  \mb_\Phi^2}{2(1-C^2)}}(x+x_0-Ct) \right] \; .
  \label{eq:nkg3}
\ee

For $\lambda <0 $ the solution is,
\be
  \Phi(x,t)=A\mb_\Phi dn\left[\beta(x+x_0-Ct),m\right] \; ,
  \label{eq:nkg4}
\ee
where $\beta=\sqrt{-\lambda \mb_\Phi^2/((1-C^2)(2-m))}$ and $A=\sqrt{2/(2-m)}$. In the limit of
$m \rightarrow 1$, one obtains the pulse solitary wave solution,
\be
  \Phi(x,t)=\mb_\Phi \sqrt{2} \mathrm{sech} \left[ \sqrt{\frac{-\lambda
  \mb_\Phi^2}{1-C^2}}(x+x_0-Ct) \right] \; .
  \label{eq:nkg5}
\ee

Therefore, one has two kind of solutions, namely the kink solution and the 
pulse solitary wave solution. These solutions can be usually used to describe
the propagation of a bubble in a torsional motion of the $Y$ model.

\subsection{Velocity as a small perturbation}

In this subsection, let us assume that the DNA-fluid interactions lead to small
perturbation of the homogeneous solution. In this case a linear perturbation
technique can be used. One can rewrite the EOM as follow,
\be
  \frac{\pd^2 \Phi}{\pd t^2}-\frac{\pd^2 \Phi}{\pd x^2}
  -m_\Phi^2 \Phi - 2g^2 v^2(x,t)\Phi + \lambda \Phi^3 =0 \; ,
  \label{eq:petur1}
\ee
According to the linear perturbation theory, the solution of Eq.
(\ref{eq:petur1}) can be treated perturbatively,
\be
   \Phi(x,t)=\Phi_0(x,t)+ \epsilon \Phi_1(x,t) \; ,
   \label{eq:petur2}
\ee 
where $\epsilon$ is a small parameter, \ie $v^2(x,t)\sim \epsilon$.
Substituting Eq. (\ref{eq:petur2}) into Eq. (\ref{eq:petur1}) leads to
the following equation,
\be
  \frac{\pd^2 \Phi_0}{\pd t^2}-\frac{\pd^2 \Phi_0}{\pd x^2}
  -m_\Phi^2 \Phi_0 + \lambda \Phi_0^3 =0 \; ,
  \label{eq:petur3}
\ee
for the zeroth order and,
\be
  \frac{\pd^2 \Phi_1}{\pd t^2}-\frac{\pd^2 \Phi_1}{\pd x^2}
  + V^2(x,t)\Phi_1 =0 \; ,
  \label{eq:petur4}
\ee
for the first order. Here $V^2(x,t)= \left(-m_\Phi^2  - 2g^2 v^2(x,t) + 3
\lambda \Phi_0^2 \right)$. Eq. (\ref{eq:petur3}) is an ordinary nonlinear
Klein-Gordon equation, while Eq. (\ref{eq:petur4}) is the linear Klein-Gordon
equation with coefficient variable.

$\Phi_0$ is a solution of Eq.(\ref{eq:petur3}) which is the term of Jacobi 
elliptic functions (Eqs. (\ref{eq:nkg2})$\sim$(\ref{eq:nkg5})). The solution is
the function of variable $x'=x-Ct$. To solve Eq. (\ref{eq:petur4}) and by
assuming that $v(x,t)=v(x')$, one needs to introduce new variables, $x'=x-Ct$
and $t'=t-Cx$ \cite{yakushevich}. In the new coordinate, Eq. (\ref{eq:petur4})
reads,
\be
  \frac{\pd^2 \Phi_1}{\pd t'^2}-\frac{\pd^2 \Phi_1}{\pd x'^2}
  + V^2(x')\Phi_1 =0 \; ,
  \label{eq:petur5}
\ee
This result yields that the solution has the form of,
\be
  \Phi(x',t')=\phi(x') \ex^{-i\omega t'} \; .
  \label{eq:petur6}
\ee
Then substituting this result into Eq. (\ref{eq:petur5}), one immediately gets 
the Sturm-Liouville equation,
\be
   -\frac{d^2 \phi}{d x'^2} + \tilde{V}^2(x') \phi =0 \; ,
   \label{eq:petur7}
\ee 
where $\tilde{V}^2(x')=V^2(x)-\omega^2$. In principles using the WKBJ
expansion technique, the solution of this equation can be obtained. One should
notice that the solution is exist only when the 'potential' slowly varies 
with $x'$. 

First, write $\tilde{V}(\epsilon x')=\tilde{V}(X)$
with  $X$ is the scale and $\epsilon \ll 1$. One has,
\be
  -\epsilon^2 \frac{d^2 \phi}{d X^2} + \tilde{V}^2(X) \phi=0 \; ,
  \label{eq:pertur8}
\ee
and the solution is \cite{dingemans},
\be
  \phi(X,\epsilon)=\frac{C_0}{\sqrt{\tilde{V}(X)}}
  \exp \left[-\frac{i}{\epsilon}\int^{x'} \tilde{V}(\xi)d\xi\right]
  +\frac{C_1}{\sqrt{\tilde{V}(X)}}\exp \left[\frac{i}{\epsilon}\int^{x'}
  \tilde{V}(\xi)d\xi\right] + \bigcirc (\epsilon) \; .
  \label{eq:pertur9}
\ee
One can also study another case with fluid velocity having the form of 
$U(x)=sin(x)$.

\subsection{Stratified velocity}

\begin{figure}[t]
\begin{center}
   \includegraphics[width=\textwidth]{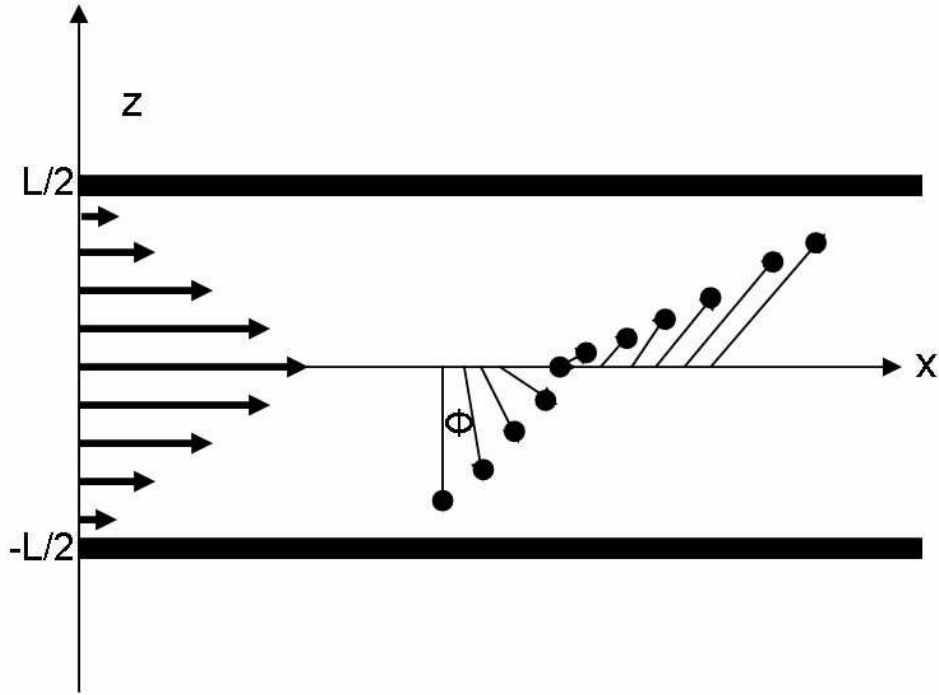}
   \caption{Geometry of torsional motion of  DNA in the stratified fluids.}
   \label{fig:strat}
\end{center}
\end{figure}

One of the important experiments is the DNA (biopolymer) dynamics in shear flow 
\cite{smith}. The experiment shows that in a steady shear flow, the molecules
continually fluctuate between nearly fully coiled and fully stretched
conformation while undergoing end-over-end tumbling. The fluctuations between
conformations are aperiodic and exhibited a power law roll-off at a high
frequency, which is attributed to a tumbling instability driven by Brownian
motion.

In the model, the calculation of shear effect on the dynamics within the $Y$
model is easy to describe. The description of such DNA bubble dynamics in shear
flow is geometrically depicted in Fig. \ref{fig:strat}.

In the microfluid experiments, the fluid is usually treated one
dimensionally and assumed to be incompressible, non-inertia, fully developed,
unidirectional, Newtonian and steady flow. Under these assumptions, the EOM for
fluid flows is \cite{smith},
\be
  0=-\frac{\pd P}{\pd x} + \frac{1}{\nu}\frac{\pd^2 v}{\pd z^2} \; ,
  \label{eq:fluid1}
\ee
where $\nu$ and $P$ are the fluid viscosity and pressure. If the pressure
gradient occurs only at one edge of channel with its length is namely $L_x$, one
can solve the equation using boundary condition $v(z=-L/2)=v(z=L/2)=0$. Then
fluid velocity in the parallel plate is given by,
\be
  v(z) =-\frac{1}{2\nu}\frac{\Delta P}{L_x} \left[ z^2 -\frac{L^2}{4}
  \right] \; .
  \label{eq:fluid2}
\ee
Note that the velocity distribution is parabolic in $z$ variable and called a
Poiseuille flow.

The EOM for the DNA dynamics is then given by,
\be
  \frac{\pd^2 \Phi}{\pd t^2}-\frac{\pd^2 \Phi}{\pd x^2}
  -U^2(z) \Phi + \lambda \Phi^3 =0 \; ,
  \label{eq:petur11}
\ee
where $U^2(z)=m_\Phi^2 + 2g^2 v^2(z)$. By solving the boundary value problem of
fluid velocity, the solution is given by Eqs. (\ref{eq:nkg3}) and
(\ref{eq:nkg5}) for $\lambda > 0$ and $\lambda <0$ respectively.

\begin{figure}[t]
        \centering
    \includegraphics[width=\textwidth]{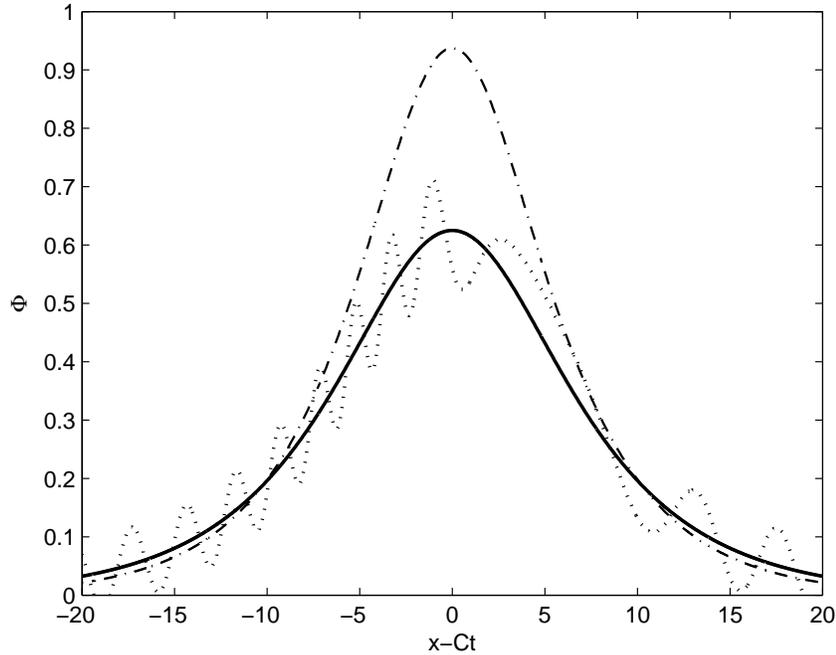}
        \caption{The rotational motion of the DNA in the term of a
        pulse solitary wave as a function of $x^\prime$ with the coupling
constants $g = 0$ (solid line) ,  $g = 0.1$ (dashed line)
        and rotational motion with small oscillation fluid velocity (dot line)
        for a fixed parameter set $(m_\Phi,v,C,\lambda) = (1,5,0.5,0.2)$.}
        \label{fig:velocity}
\end{figure}

\begin{figure}[t]
        \centering
    \includegraphics[width=8cm,angle=0,trim=0 0 0 0]{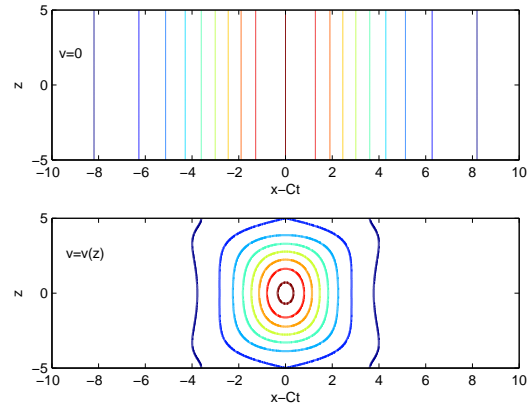}
        \caption{The rotational motion of  DNA in  term of a
        pulse solitary wave in the Poiseuille flow with a fixed parameter set
$(m_\Phi,v,C,\lambda) = (1,5,0.5,0.2)$}
        \label{fig:stratif}
\end{figure}

\section{Summary}

A new model for DNA dynamics immersed in fluid system has been introduced. The
model is constructed within the gauge field theory framework to describe the
interactions between DNA matter and surrounding fluid from the first principle.
The dynamics is then studied for various types of fluid velocity.

The basic behavior of the bubble dynamics is described by a pulse solitary
waves in Eq. (\ref{eq:nkg5}). The open state dynamics, that is usually 
described by the soliton solution, in the present model is realized in the
case of uniform fluid velocity as depicted in Fig. \ref{fig:velocity}.
The open bubble should correspond to the nonlinear excitation, and thus it is
induced by the nonlinear dynamics of DNA double helix itself \cite{cadoni}.
From the figure, one can conclude that the bubble amplitude is getting greater
as the fluid flows at a constant velocity. On the other hand, the dotted line
represents the effects of torsional motion on the DNA immersed in the
oscillating fluid, \ie $v=sin(\omega x)$. However, the oscillation in the bubble
dynamics is relatively small since by assumption the perturbation term (fluid
velocity) is kept small.

Another effects due to stratified flow in the bubble dynamics is shown in Fig.
\ref{fig:stratif}. As mentioned above, according to the experiments the average
extension of DNA under shear flow was found to increase gradually with
increasing velocity gradient, reaching a plateau below half-full
extension. The hydrodynamics friction is more important factor to aligned of the
DNA molecule in a direction parallel to the flow. The present result is
given in Fig. \ref{fig:stratif} which shows that the fluid flow increases the
soliton amplitude with the same profile, that is it reaches the peak at the
center and almost unaffected near the plateau.

Further, the effects of external forces like electromagnetic force might be
interesting to be investigated. This might be important as one considers the
nanofluidic system \cite{viovy}. This will be worked out and discussed in the
next-coming works.

\section*{Acknowledgments}

AS thanks the DAP Consultant for partial funding and the Group for Theoretical
and Computational Physics LIPI for warm hospitality during the work. This work
is funded by the Indonesia Ministry of Research and Technology and the Riset
Kompetitif LIPI in fiscal year 2011 under Contract no.  11.04/SK/KPPI/II/2011.

\bibliographystyle{ws-procs975x65}
\bibliography{sulaiman}

\end{document}